\documentclass[11pt]{article}
\usepackage{geometry}                
\usepackage[parfill]{parskip}
\usepackage{amssymb}
\usepackage{graphicx}
\usepackage{amsmath}

\title{Green's functions for reflectionless potentials and addition of boundstates to powerlaw potentials to find Supersymmetric partners}

\author{C. V. Sukumar \\{\em Wadham College,}\\{\em University of Oxford, Oxford OX1 3PN, U.K.}}

\begin{document}

\maketitle

\begin{abstract}
Green's functions for reflectionless potentials are constructed and analyzed. Green's functions for power law potentials, their Super Symmetric partners and sum rules for eigenvalues are examined. The SUSY partner potentials to power law potentials which have an additional bound state at $E=0$ are constructed.
\end{abstract}

\section{Green's functions}

Reflectionless potentials can belong to one of two categories- either (1)- potentials which support negative energy eigenstates for a Schrodinger differential operator and also have a continuous spectrum with vanishing reflection coefficients for positive energies or (2)-potentials which rise to infinite values and have a only discrete spectrum with positive energy eigenvalues.
 
\subsection{Confining potentials}

We first consider confining potentials which have only a bound-state spectrum with orthonormal eigenfunctions  $\psi_j$ and positive eigenvalues $\gamma_j^2$ for the Schrodinger equation for a potential $U(x)$ in a domain $[a,b]$ using units in which $\hbar=1$ and mass $m=1/2$:
\begin{equation}
\frac{d^2}{dx^2}\psi_j(x) = \big(U-\gamma_j^2\big) \psi_j(x) \label{eq:d1}
\end{equation}
The orthonormal condition and the completeness relation are
\begin{align}
\int_{a}^{b} \psi_j(x)\ \psi_k(x)\ dx &= \delta_{jk} \label{}\\
g_0(x,x^{\prime}) &\equiv\sum_j \psi_j(x) \psi_j(x^{\prime}) = \delta(x-x^{\prime}) \label{eq:c0}
\end{align}

Using (\ref{eq:d1}) it can be established that
\begin{align}
g_k(x,x^{\prime}) &\equiv \sum_j  (\gamma_j)^k \psi_j(x) \psi_j((x^{\prime}) ,\ k=0,\pm 1, \pm 2, \ \dots \label{eq:d0}\\
\frac{d^2}{dx^2} g_k(x,x^{\prime}) &=U g_k(x,x^{\prime}) - g_{k+2}(x,x^{\prime}) \label{}\\
g_{k+n}(x,x^{\prime}) &= \int_{a}^{b}g_k(x,z) g_n(z,x^{\prime})\ dz \label{eq:p1}
\end{align}
The Green's function $G(x,x^{\prime})$ for the Schrodinger equation may be identified as corresponding to $-g_{-2}(x,x^{\prime})$ and 
\begin{equation}
G(x,x^{\prime}) \equiv - g_{-2}(x,x^{\prime})\quad \Rightarrow\ \Big[\frac{d^2}{dx^2} - U\Big]G(x,x^{\prime})=\delta(x-x^{\prime}) \label{eq:g1}
\end{equation}

Using (\ref{eq:d1}) it can also be established that the squares of the eigenfunctions satisfy
\begin{equation}
\frac{d^3}{dx^3}\psi_j^2(x) = 8[U-\gamma_j^2]\psi_j \frac{d}{dx}\psi_j +2\psi_j^2  \frac{du}{dx}\label{}
\end{equation}
It follows that
\begin{equation}
\big[\frac{d^3}{dx^3} - 4U \frac{d}{dx}-2 \frac{du}{dx}\big] g_k(x,x)= -4\frac{d}{dx} g_{k+2}(x,x) \label{eq:d2}
\end{equation}
which can also be written in the integral form
\begin{equation}
g_{k+2}(x,x)= \frac{1}{4} \Big(\Big[\big(-\frac{d^2}{dx^2} + 4U\big) g_k(x,x)\Big] \vert_{a}^x - 2\int_{a}^x \frac{dU}{dy} g_k(y,y) dy\Big)\label{eq:d3}
\end{equation}
Knowledge of $g_1(x,x)$ enables the determination of $g_{2k+1}$ through the repeated application of the integral form in (\ref{eq:d3}). Knowledge of $g_{-2}(x,x)$ enables the determination of $g_{-2k-2}(x,x)$ by solving the differential equation in (\ref{eq:d2}). Two boundary conditions on the solutions of the Schrodinger equation generate solutions for all $x$ and hence give conditions on all the derivatives of the eigenstates and consequently generate boundary conditions on all the derivatives of $g_k$ guaranteeing that the third order differential equation has adequate boundary conditions to generate unique solutions. It can be seen that 
\begin{equation}
\int_{a}^{b} g_k(x,x) dx = \sum_j \big(\gamma_j\big)^k \label{eq:d4}
\end{equation}
which is a set of sum rules for the eigenvalues of the Schrodinger equation for a confining potential. The sums for negative values of $k$ may converge and the sums for positive values of $k$ will diverge.

${\mathbf{Theorem}}$: It can be seen that
\begin{align}
F(x,x^{\prime})&\equiv f(x_<)g(x_>)\ \Rightarrow \label{}\\
\int_a^b \int_a^b F(x,x^{\prime}) dx dx^{\prime}&= \int_a^b g(x)\Big[\int_a^x f(x^{\prime}) dx^{\prime}\Big] dx + \int_a^b f(x)\Big[\int_x^b g(x^{\prime}) dx^{\prime}\Big] dx \label{eq:Th1}
\end{align}
where $x_<$ is the lesser of $(x,x^{\prime})$ and $x_>$ is the greater of $(x,x^{\prime})$. By changing the order of the $x$ and $x^{\prime}$ integrations in the second integral we get
\begin{align}
\int_a^b \int_a^b F(x,x^{\prime}) dx dx^{\prime}&= \int_a^b g(x)\Big[\int_a^x f(x^{\prime}) dx^{\prime}\Big] dx + \int_a^b g(x^{\prime})\Big[\int_ax^{\prime} f(x) dx\Big] dx^{\prime} \label{}\\
&= 2\int_a^b g(x)\Big[\int_a^x f(x^{\prime}) dx^{\prime}\Big] dx =2\int_a^b f(x)\Big[\int_x^b g(x^{\prime}) dx^{\prime}\Big] dx \label{eq:Th2}
\end{align}
The equivalence expressed in (\ref{eq:Th1}) and (\ref{eq:Th2}) may be used together with the associative property expressed in (\ref{eq:p1}) to evaluate the sum rules in (\ref{eq:d4})

\subsection{reflectionless potentials related to KdV solitons}

The discussion for confining potentials can be adapted for reflectionless potentials with a discrete negative energy spectrum which may be related to multi-soilton solutions of the KdV equation (Scott {\it et al} 1973).
The mapping 
\begin{equation}
\gamma_j\ \Rightarrow\ i \alpha_j ,\  L_k(x)\equiv -2(-2i)^{2k+1} g_{2k+1}(x,x)=-2\sum_j\big(2\alpha_j\big)^{2k+1}\psi^2_j(x) \label{eq:L1}
\end{equation} 
may be used to establish that the KdV equation is the $k=0$ member of the Lax hierarchy (Lax 1968, Gardner{\it et al} 1967) $L_k$
\begin{equation}
L_0(x)= -4\sum_j \alpha_j \psi^2_j(x) = U(x) \quad \Rightarrow \  \int_{-\infty}^{\infty} U(x) dx = -4\sum_j \alpha_j \label{eq:L2}
\end{equation}
where $\psi_j$ are the bound-state solutions of the reflectionless potential $U$ at energies $-\alpha_j^2$. $U$ can be interpreted in terms of soliton solutions whose time evolution is governed by the KdV equation.

All the odd members of the $g_k(x,x)$ hierarchy can be related to members of the Lax hierarchy as defined in (\ref{eq:L1}) and the time evolution of the soliton solutions of the Lax hierarchy is governed by:
\begin{align}
\frac{\partial U}{\partial t_k} &= -\frac{\partial L_k}{\partial x} \label{eq:L3}\\
L_k&=-2 \sum_j \big[2\alpha_j\big]^{2k+1} \psi^2_j(x) \label{eq:L4}\\
\frac{d}{dx} L_{k+1} &=\Big[\frac{d^3}{dx^3} - 4U \frac{d}{dx}-2 \frac{du}{dx}\Big] L_k \label{eq:L5}
\end{align}
leading to the sum rules
\begin{equation}
\int_{-\infty}^{\infty} L_k(x) dx = -2\sum_j \big[2\alpha_j\big]^{2k+1} \label{eq:L6}
\end{equation}
The sums may converge for positive values of $k$ and may diverge for negative values of $k$.

\section{Example - Particle in a box}

\subsection{Case 1}
Consider the solutions for a free particle confined within a box with infinite walls at $x=0$ and $x=1$. The eigenfunctions which satisfy $\psi_j(0)=0$ and $\psi_j(1)=0$ are
\begin{align}
\psi_j(x)&= \sqrt{2} \sin(\gamma_j x) ,\quad \gamma_j = j\pi \Rightarrow \label{eq:d7}\\
g_{-1}(x,x^{\prime})&= 2 \sum_{j=1}^{\infty} \frac{\sin(j\pi x) \sin(j\pi x^{\prime})}{j\pi}=\sum_{j=1}^{\infty} \frac{\big(\cos[j\pi (x-x^{\prime})]-\cos[j\pi(x+x^{\prime})]\big)}{j\pi} \notag \\
&= \frac{1}{\pi} \ln\Big(\frac{\sin[(x+x^{\prime})\pi/2]}{\sin[|(x-x^{\prime})|\pi/2]}\Big) \label{}\\
g_{-2}(x,x^{\prime})&= 2 \sum_{j=1}^{\infty} \frac{\sin(j\pi x) \sin(j\pi x^{\prime})}{j^2\pi^2} =\sum_{j=1}^{\infty} \frac{\big(\cos[j\pi (x-x^{\prime})]-\cos[j\pi(x+x^{\prime})]\big)}{j^2\pi^2} \notag\\
&= \Big[\frac{1}{6}-\frac{|(x-x^{\prime})|}{2}+\frac{\big(x-x^{\prime}\big)^2}{4}\Big]-\Big[\frac{1}{6}-\frac{(x+x^{\prime})}{2}+\frac{\big(x+x^{\prime}\big)^2}{4}\Big]\label{}\\
&=x_< (1-x_>)\label{eq:g2}
\end{align}
where $x_<$ is the lesser of $(x,x^{\prime})$ and $x_>$ is the greater of $(x,x^{\prime})$. The two results for the two series given above are well known (Gradshteyn and Ryzhik 1.441.2 p38, 1.443.3 p39). Application of the result in (\ref{eq:p1}) for $n=-1$ and $k=-1$ leads to the relation
\begin{equation}
\frac{1}{\pi^2} \int_0^1 \ln\Big(\frac{\sin[(x+z)\pi/2]}{\sin[|(x-z)|\pi/2]}\Big) \ln\Big(\frac{\sin[(z+x^{\prime})\pi/2]}{\sin[|(z-x^{\prime})|\pi/2]}\Big) dz=x_< (1-x_>)\label{eq:q1}
\end{equation}
It can also be verified that
\begin{equation}
\int_0^1 g_{-2}(x,x) dx = \int_0^1 x(1-x) dx = \frac{1}{6} =\sum_{j=1}^{\infty} \frac{1}{j^2\pi^2} \label{}
\end{equation}

It was noted earlier that $-g_{-2}(x,x^{\prime})$ can be identified with the Green's function $G(x,x^{\prime})$ which satisfies (\ref{eq:g1}). $G(x,x^{\prime})$ can also be constructed by an independent procedure by finding the solutions for eigenvalue $0$ which satisfy the appropriate boundary condition at $x=0$ and $x=1$. For the particle in a box the appropriate zero energy solutions are $\psi=x$ and $\psi=(1-x)$ and
\begin{equation}
G(x,x^{\prime})= -x_< \big(1-x_>\big)\ \Rightarrow\ G(x,x^{\prime})=x(x^{\prime}-1)\Theta(x^{\prime}-x) + x^{\prime}(x-1)\Theta(x-x^{\prime}) \label{eq:g3}
\end{equation}
where $\Theta(z)$ is the Heaviside step function which vanishes for negative values of $z$ and has value 1 for positive $z$. $G$ so defined satisfies (\ref{eq:g1}).

The knowledge of $G(x,x^{\prime})$ enables the identification of two possible ways of finding $g_{-4}(x,x)$ either by solving (\ref{eq:d2}) using $g_{-2}(x,x)$ as input or by using (\ref{eq:p1}) with $k=n=-2$. The second method yields
\begin{align}
g_{-4}(x,x)&=\int_0^1 G(x,x^{\prime}) G(x^{\prime},x) dx^{\prime} = \int_0^x (1-x)^2 x^{\prime2} dx^{\prime} +\int_x^1 x^2(1-x^{\prime})^2 dx^{\prime} \notag\\
&=\frac{x^2}{3}(1-x)^2\label{eq:g4}\\
\int_0^1g_{-4}(x,x)dx&= \frac{1}{90} =\sum_{j=1}^{\infty} \frac{1}{j^4\pi^4} \label{eq:g5}
\end{align}
The alternate procedure of solving using (\ref{eq:d2}) is to integrate both sides of the equation in the range $[0,x]$ three times to find:
\begin{align}
\Big[\frac{d^2}{dx^2} g_{-4}(x,x)\Big]\vert_0^x&= -4\Big[g_{-2}(x,x)\Big]\vert_0^x=-4x(1-x) \label{}\\
\Big[\frac{d}{dx} g_{-4}(x,x)\Big]\vert_0^x&=x\Big[\frac{d^2}{dx^2} g_{-4}(x,x)\Big]\vert_0 -2x^2+\frac{4x^3}{3}\label{}\\
g_{-4}(x,x)&=x\Big[\frac{d}{dx} g_{-4}(x,x)\Big]\vert_0+\frac{x^2}{2}\Big[\frac{d^2}{dx^2} g_{-4}(x,x)\Big]\vert_0 -\frac{2x^3}{3} + \frac{x^4}{3}\label{}
\end{align}
The boundary condition on $\psi_j$ shows that the derivative of $g_{-4}$ at $x=0$ is $0$. The second derivative of $g_{-4}$ at $x=0$ may be determined to have the value $2/3$ by imposing the condition that $g_{-4}$ must vanish at $x=1$. These manipulations lead to 
\begin{equation}
g_{-4}(x,x)= \frac{1}{3} (x^4-2x^3+x^2) = \frac{1}{3} x^2 (1-x)^2 \label{}
\end{equation}
in agreement with (\ref{eq:g4}).

\subsection{Case 2}

If we consider the eigenfunctions for a free particle which satisfy $[d\psi_j(x)/dx]=0$ at $x=0$ and $\psi_j(1)=0$
\begin{align}
\psi_j(x)&= \sqrt{2} \cos(\gamma_j x) ,\quad \gamma_j = (j-1/2)\pi \Rightarrow \label{eq:d8}\\
g_{-1}(x,x^{\prime})&= 2 \sum_{j=1}^{\infty} \frac{1}{\gamma_j}\cos(\gamma_j x) \cos(\gamma_j x^{\prime})=\sum_{j=1}^{\infty} \frac{1}{\gamma_j}\big(\cos[\gamma_j (x+x^{\prime})] +\cos[\gamma_j (x-x^{\prime})]\big) \notag\\
&= \frac{1}{\pi} \ln\Big(\cot[(x+x^{\prime})\pi/4] \cot[|(x-x^{\prime})|\pi/4]\Big) \label{}\\
g_{-2}(x,x^{\prime})&= 2 \sum_{j=1}^{\infty} \frac{1}{\gamma_j^2}\cos(\gamma_j x)\cos(\gamma_j y)= \sum_{j=1}^{\infty} \frac{1}{\gamma_j^2}\big(\cos[\gamma_j (x+y)] +\cos[\gamma_j (x-y)\big)\notag\\
&=\frac{1}{2}\Big[1-(x+x^{\prime})\Big]+ \frac{1}{2}\Big[1-|(x-x^{\prime})|\Big] =(1-x_>)\label{eq:g6}
\end{align}
where $x_>$ is the greater of $(x,x^{\prime})$. The two results for the two series given above are well known (Gradshteyn and Ryzhik 1.442.2 p38, 1.444.6 p39). Application of the result in (\ref{eq:p1}) for $n=-1$ and $k=-1$ leads to the relation
\begin{align}
&\frac{1}{\pi^2} \int_0^1 \ln\Big(\cot[(x+z)\pi/4] \cot[|(x-z)|\pi/4\Big)\ \ln\Big(\cot[(z+x^{\prime}\pi/4)]\cot[(|z-x^{\prime}|)\pi/4]\Big) dz\notag\\
&= (1-x_>)\label{eq:q2}
\end{align}
It can also be verified that
\begin{equation}
\int_0^1 g_{-2}(x,x) dx = \int_0^1 (1-x) dx = \frac{1}{2} =\sum_{j=1}^{\infty} \frac{4}{(2j-1)^2\pi^2} \label{}
\end{equation}

As in the earlier case $-g_{-2}(x,x^{\prime})$ can be identified with the Green's function $G(x,x^{\prime})$ which satisfies (\ref{eq:g1}). For a free particle the zero energy solutions are $\psi=1$ which has vanishing derivative at $x=0$ and $\psi=(1-x)$ which vanishes at $x=1$ and these solutions may be used to construct $G(x,x^{\prime})$ :
\begin{equation}
G(x,x^{\prime})= -\big(1-x_>\big)\ \Rightarrow\ G(x,x^{\prime})=(x^{\prime}-1)\Theta(x^{\prime}-x) + (x-1)\Theta(x-x^{\prime}) \label{eq:g7}
\end{equation}
which satisfies (\ref{eq:g1}). The knowledge of $G(x,x^{\prime})$ enables the identification of two possible ways of finding $g_{-4}(x,x)$ either by solving (\ref{eq:d2}) using $g_{-2}(x,x)$ as input or by using (\ref{eq:p1}) with $k=n=-2$. The second method yields
\begin{align}
g_{-4}(x,x)&=\int_0^1 G(x,x^{\prime}) G(x^{\prime},x) dx^{\prime} = \int_0^x (1-x)^2  d x^{\prime}+\int_x^1 (1-x^{\prime})^2 dx^{\prime} \notag\\
&= x(1-x)^2 +\frac{(1-x)^3}{3} = \frac{1}{3}(1-x)^2 (2x+1)\label{eq:g8}\\
\int_0^1g_{-4}(x,x)dx&= \frac{1}{6} =\sum_{j=1}^{\infty} \frac{16}{(2j+1)^4\pi^4} \label{eq:g9}
\end{align}
The alternate procedure of solving using (\ref{eq:d2}) is to integrate both sides of the equation in the range $[x,1]$ three times to find:
\begin{align}
\Big[\frac{d^2}{dx^2} g_{-4}(x,x)\Big]\vert_1^x&= -4\Big[g_{-2}(x,x)\Big]\vert_1^x=-4(1-x) \label{}\\
\Big[\frac{d}{dx} g_{-4}(x,x)\Big]\vert_1^x&=(x-1)\Big[\frac{d^2}{dx^2} g_{-4}(x,x)\Big]\vert_1 -4x+2x^2+2\label{}\\
g_{-4}(x,x)&=(x-1)\big[\frac{d}{dx} g_{-4}(x,x)\big]\vert_1+\frac{(x-1)^2}{2}\Big[\frac{d^2}{dx^2} g_{-4}(x,x)\Big]\vert_1 - \notag\\
& -2x^2+\frac{2x^3}{3} +2x - \frac{2}{3}\label{}
\end{align}
The boundary condition on $\psi_j$ shows that the first derivative of $g_{-4}$ vanishes at $x=0$ and $x=1$. The second derivative of $g_{-4}$ at $x=1$ may be determined to have the value 2 by imposing the condition that the derivative of $g_{-4}$ must vanish at $x=0$. These manipulations lead to 
\begin{equation}
g_{-4}(x,x)= \frac{1}{3} (2x^3-3x^2+1) = \frac{1}{3}(1-x)^2(2x+1)\label{}
\end{equation}
in agreement with (\ref{eq:g8}).

\subsection{Case 3}

If we consider the eigenfunctions for a free particle which satisfy $[d\psi_j(x)/dx]=0$ at $x=1$ and $\psi_j(0)=0$
\begin{align}
\psi_j(x)&= \sqrt{2} \sin(\gamma_j x) ,\quad \gamma_j = (j-1/2)\pi \Rightarrow \label{eq:d9}\\
g_{-1}(x,x^{\prime})&= 2 \sum_{j=1}^{\infty} \frac{1}{\gamma_j}\sin(\gamma_j x) \sin(\gamma_j x^{\prime})=\sum_{j=1}^{\infty} \frac{1}{\gamma_j}\big(\cos[\gamma_j (x-x^{\prime})] -\cos[\gamma_j (x+x^{\prime})]\big) \notag\\
&= \frac{1}{\pi} \Big[\ln\Big(\cot[|(x-x^{\prime})|\pi/4]\Big)-\ln\Big(\cot[(x+x^{\prime})\pi/4]\Big)\Big] \label{}\\
g_{-2}(x,x^{\prime})&= 2 \sum_{j=1}^{\infty} \frac{1}{\gamma_j^2}\sin(\gamma_j x)\sin(\gamma_j x^{\prime})= \sum_{j=1}^{\infty} \frac{1}{\gamma_j^2}\big(\cos[\gamma_j (x-x^{\prime})] -\cos[\gamma_j (x+x^{\prime})]\big)\notag\\
&= \frac{1}{2} \Big(\big[1-|(x-x^{\prime})|\big] - \big[1- (x+x^{\prime})\big]\Big)= x_< \label{eq:g11}\\
G(x,x^{\prime})&= -x_<=-x^{\prime} \Theta(x-x^{\prime}) -x \Theta(x^{\prime}-x)\label{}
\end{align}
which satisfies (\ref{eq:g1}). These identifications lead to the relation
\begin{equation}
\frac{4}{\pi^2} \int_0^1 \Big[\ln\frac{\Big(\cot(|[x-z]|\pi/4)\Big)}{\Big( \cot([x+z]\pi/4)\Big)}\Big]\Big[\ln\frac{\Big(\cot(|[z-x^{\prime}]|\pi/4)\Big)}{\Big( \cot([z+x^{\prime}]\pi/4)\Big)}\Big]   dz = x_<\label{eq:q3}
\end{equation}
It can also be verified that
\begin{equation}
\int_0^1 g_{-2}(x,x) dx = \int_0^1 x dx = \frac{1}{2} =\sum_{j=1}^{\infty} \frac{4}{(2j-1)^2\pi^2} \label{}
\end{equation}
It can also be established that
\begin{align}
g_{-4}(x,x)&=\int_0^1 G(x,x^{\prime}) G(x^{\prime},x) dx^{\prime} = \int_0^x  x^{\prime2} dx^{\prime} +\int_x^1 x^2\ dx^{\prime} = x^2-\frac{2x^3}{3} \label{eq:g12}\\
\int_0^1g_{-4}(x,x)dx&= \frac{1}{6} = \sum_{j=1}^{\infty} \frac{16}{(2j+1)^4\pi^4} \label{eq:g13}
\end{align}

\subsection{Case 4}
If we consider the eigenfunctions for a free particle which satisfy $[d\psi_j(x)/dx]=0$ at $x=0$ and $x=1$
\begin{align}
\psi_j(x)&= \sqrt{2} \cos(\gamma_j x) ,\quad \gamma_j = j\pi \Rightarrow \label{eq:d12}\\
g_{-1}(x,x^{\prime})&= 2 \sum_{j=1}^{\infty} \frac{\cos(j\pi x) \cos(j\pi x^{\prime})}{j\pi}=\sum_{j=1}^{\infty} \frac{1\big(\cos[j\pi (x-x^{\prime})]+\cos[j\pi(x+x^{\prime})]\big)}{j\pi} \notag \\
&= \frac{-1}{\pi} \ln\Big(4\sin[|(x-x^{\prime})|\pi/2] \sin[(x+x^{\prime})\pi/2]\Big) \label{}\\
g_{-2}(x,x^{\prime})&= 2 \sum_{j=1}^{\infty} \frac{\cos(j\pi x) \cos(j\pi x^{\prime})}{j^2\pi^2} =\sum_{j=1}^{\infty} \frac{1\big(\cos[j\pi (x-x^{\prime})]+\cos[j\pi(x+x^{\prime})]\big)}{j^2\pi^2} \notag\\
&=\Big(\Big[\frac{1}{6}-\frac{|(x-x^{\prime})|}{2} +\frac{\big(x-x^{\prime}\big)^2}{4}\Big] +\Big[\frac{1}{6}-\frac{(x+x^{\prime})}{2} +\frac{(x+x^{\prime})^2}{4}\Big]\Big)\notag\\
&=\Big(\frac{1}{3}-x_> +\frac{1}{2}(x^2+x^{\prime2})\Big)=-G(x,x^{\prime})\label{eq:g15}
\end{align}
These identifications lead to the relation
\begin{align}
&\frac{1}{\pi^2} \int_0^1 \ln\Big(4\sin(|[x-z]|\pi/2)\sin([x+z]\pi/2)\Big) \ln\Big(4\sin(|[z-x^{\prime}]|\pi/2)\sin([z+x^{\prime}]\pi/2)\Big) dz \notag\\
&=\Big(\frac{1}{3}-x_> +\frac{1}{2}\big(x^2+x^{\prime2}\big)\Big)\label{eq:q4}
\end{align}
It can also be verified that
\begin{equation}
\int_0^1 g_{-2}(x,x) dx = \int_0^1\Big(\frac{1}{3}-x+x^2\Big)  dx = \frac{1}{6} =\sum_{j=1}^{\infty} \frac{1}{j^2\pi^2} \label{}
\end{equation}

The completeness relation defined in (\ref{eq:c0}) should also include the $j=0$ term in the sum for this case because the normalized $j=0$ function defined by (\ref{eq:d12}) is $\psi_0\rightarrow1$ and is nonvanishing. But the $g_k$ functions defined above for $k\ne0$ have the sums beginning with $j=1$. Hence it can be recognized that in this case 
\begin{equation}
G(x,x^{\prime})=\Big(-\frac{1}{3}+x_>-\frac{1}{2}(x^2+x^{\prime2})\Big) =-\Big(\frac{1}{3} +\frac{1}{2}(x^2+x^{\prime2})\Big)+x\Theta(x-x^{\prime})+x^{\prime}\Theta(x^{\prime}-x)\label{}
\end{equation}
will satisfy
\begin{equation}
\Big[\frac{d^2}{dx^2} - U\Big]G(x,x^{\prime})=\delta(x-x^{\prime}) -1  \label{eq:ga1}
\end{equation}
instead of (\ref{eq:g1})

It can be established that
\begin{align}
g_{-4}(x,x)&=\int_0^1 G(x,y) G(y,x) dy \notag\\
&=\int_0^x \Big(\frac{1}{3}-x +\frac{1}{2}(x^2+y^2)\Big)^2  dy +\int_x^1\Big(\frac{1}{3}-y +\frac{1}{2}(x^2+y^2)\Big)^2  dy \label{}\\
&=\frac{(-21x^5+45x^4-35x^3+15x^2-5x+1)}{45} + \notag{}\\
&+\quad\quad \frac{(21x^5-60x^4+65x^3-30x^2+5x)}{45}\label{}\\
&=\frac{ (-15x^4+30x^3-15x^2+1)}{45} =\frac{1}{45} -\frac{1}{3}x^2(1-x)^2 \label{eq:g16}\\
\int_0^1g_{-4}(x,x)dx&= \frac{1}{90} = \sum_{j=1}^{\infty} \frac{1}{j^4\pi^4} \label{eq:g17}
\end{align}

\subsection{ Series with alternating signs}

We can also consider weighted eigenstate sums with alternating signs. If we define
\begin{equation}
f_k(x,x^{\prime}) \equiv \sum_{j=1}  (-)^{j-1}(\gamma_j)^k \psi_j(x) \psi_j(x^{\prime}) ,\ k=0,\pm 1, \pm 2, \ \dots \label{}
\end{equation}
then the $f_k$ are related to the $g_k$ defined earlier by the relation
\begin{equation}
g_{k+n}(x,x^{\prime}) = \int_{a}^{b}f_k(x,z) f_n(z,x^{\prime})\ dz \label{eq:p2}
\end{equation}
We can consider each of the 4 cases considered in the previous section with alternating signs for the series.

$\bullet$
For the particle in a box eigenstates which satisfy $\psi_j(0)=0$ and $\psi_j(1)=0$ it can be shown using (\ref{eq:d7}) that (Gradshteyn and Ryzhik 1.441.4 p38)
\begin{align}
\psi_j(x)&= \sqrt{2} \sin(\gamma_j x) ,\quad \gamma_j = j\pi \Rightarrow \notag\\
f_{-1}(x,x^{\prime})&= 2 \sum_{j=1}^{\infty} \frac{(-)^{j-1}}{j\pi}\sin(j\pi x) \sin(j\pi x^{\prime})\notag\\
&=\sum_{j=1}^{\infty} \frac{(-)^{j-1}}{j\pi}\big(\cos[j\pi (x-x^{\prime})]-\cos[j\pi(x+x^{\prime})]\big) \notag \\
&= \frac{1}{\pi} \ln\Big(\frac{\cos[(x-x^{\prime})\pi/2]}{\cos[(x+x^{\prime})\pi/2]}\Big) \label{}
\end{align}
which leads to a relation similar to (\ref{eq:q1}) of the form
\begin{align}
g_{-2}(x,x^{\prime})&= \int_0^1 f_{-1}(x,z) f_{-1}(z,x^{\prime}) dz \ \Rightarrow \notag\\
x_< (1-x_>)&=\frac{1}{\pi^2} \int_0^1 \ln\Big(\frac{\cos([x-z]\pi/2)}{\cos([x+z]\pi/2)}\Big)\ \ln\Big(\frac{\cos([z-x^{\prime}]\pi/2)}{\cos([z+x^{\prime}]\pi/2)}\Big) dz\label{eq:q5}
\end{align}

$\bullet$
Similarly for eigenstates which satisfy $[d\psi_j(x)/dx]=0$ at $x=0$ and $\psi_j(x)=0$ at $x=1$ it can be shown using (\ref{eq:d8}) that (Gradshteyn Ryzhik 1.442.4 p38)
\begin{align}
\psi_j(x)&= \sqrt{2} \cos(\gamma_j x) ,\quad \gamma_j = (j-1/2)\pi \Rightarrow \notag\\
f_{-1}(x,x^{\prime})&= 2 \sum_{j=1}^{\infty} \frac{(-)^{j-1}}{\gamma_j\pi}\cos(\gamma_j x) \cos(\gamma_j x^{\prime})\notag\\
&= \sum_{j=1}^{\infty} \frac{(-)^{j-1}}{\gamma_j}\big(\cos[\gamma_j (x+x^{\prime})] +\cos[\gamma_j (x-x^{\prime})]\big) \label{}
\end{align}
The Fourier analysis of the function $[\Theta(1-x)\ - 1/2]$ in the range $[0\le x\le 2]$ in terms of even functions gives 
\begin{align}
\sum_{j=1}^{\infty} (-)^{j-1} \frac{\cos[(j-1/2)\pi x]}{(j-1/2)}&= +\frac{1}{2}\ \hbox{if}\ 0\le x< 1 \label{eq:t1}\\
&= -\frac{1}{2}\ \hbox{if}\ 1 < x\le 2 \label{eq:t2}
\end{align}
Using this result it can be shown that
\begin{align}
f_{-1}(x,x^{\prime})&= 1\ \hbox{if}\ x+x^{\prime} \le 1\ ,\ =0\ \hbox{if}\ x+ x^{\prime}>1 \ \Rightarrow \label{}\\
f_{-1}(x,x^{\prime})&= \Theta(1-x-x^{\prime}) \label{}
\end{align}
which leads to a relation similar to (\ref{eq:q2}) of the form
\begin{equation}
g_{-2}(x,x^{\prime}) = \int_0^1 f_{-1}(x,z) f_{-1}(z,x^{\prime}) dz= \int_0^{1-z_>} dz = 1-x_> \label{eq:q6}
\end{equation}
where $x_>$ is the greater of $(x,x^{\prime})$.

$\bullet$
For eigenfunctions which satisfy $[d\psi_j(x)/dx]=0$ at $x=1$ and $\psi_j(0)=0$ it can be shown using (\ref{eq:d9}) that
\begin{align}
\psi_j(x)&= \sqrt{2} \sin(\gamma_j x) ,\quad \gamma_j = (j-1/2)\pi \Rightarrow \notag\\
f_{-1}(x,x^{\prime})&= 2 \sum_{j=1}^{\infty} \frac{(-)^{j-1}}{\gamma_j}\sin(\gamma_j x) \sin(\gamma_j x^{\prime})\notag\\
&= \sum_{j=1}^{\infty} \frac{(-)^{j-1}}{\gamma_j}\big(\cos[\gamma_j (x-x^{\prime})] -\cos[\gamma_j (x+x^{\prime})]\big) \label{}
\end{align}
Using the result given in (\ref{eq:t1}) and (\ref{eq:t2}) it can be shown that 
\begin{align}
f_{-1}(x,x^{\prime})&= 1\ \hbox{if}\ x+x^{\prime} \ge 1\ ,\ =0\ \hbox{if}\ x+ x^{\prime}<1 \ \Rightarrow \label{}\\
f_{-1}(x,x^{\prime})&= \Theta(x+x^{\prime}-1) \label{}
\end{align}
which leads to a relation similar to (\ref{eq:q3}) of the form
\begin{equation}
g_{-2}(x,x^{\prime}) = \int_0^1 f_{-1}(x,z) f_{-1}(z,x^{\prime}) dz= \int_{1-x_<}^1 dz = x_< \label{eq:q7}
\end{equation}
where $x_<$ is the lesser of $(x,x^{\prime})$.

$\bullet$
For eigenfunctions which satisfy $[d\psi_j(x)/dx]=0$ at $x=0$ and $x=1$ it can be shown using (\ref{eq:d12}) that
\begin{align}
\psi_j(x)&= \sqrt{2} \cos(\gamma_j x) ,\quad \gamma_j = j\pi \Rightarrow \notag\\
f_{-1}(x,x^{\prime})&= 2 \sum_{j=1}^{\infty} \frac{(-)^{j-1}}{j\pi}\cos(j\pi x) \cos(j\pi x^{\prime})\notag\\
&=\sum_{j=1}^{\infty} \frac{(-)^{j-1}}{j\pi}\big(\cos[j\pi (x+x^{\prime})]-\cos[j\pi(x-x^{\prime})]\big) \notag \\
&= \frac{1}{\pi} \ln\Big(4\cos[(x-x^{\prime})\pi/2] \cos[(x+x^{\prime})\pi/2]\Big) \label{}
\end{align}
which leads to an identity similar to (\ref{eq:q4}) in the form
\begin{align}
&g_{-2}(x,x^{\prime})= \int_0^1 f_{-1}(x,z) f_{-1}(z,x^{\prime}) dz \ \Rightarrow \notag\\
&=\int_0^1 \frac{\ln\Big(4\cos[(x-z)\pi/2] \cos[(x+z)\pi/2]\Big)}{\pi} \frac{\ln\Big(4\cos[(z-x^{\prime})\pi/2] \cos[(z+x^{\prime})\pi/2]\Big)}{\pi} dz\notag\\
&=\Big(\frac{1}{3}-x_>+\frac{1}{2}(x^2+x^{\prime2}) \Big)\label{eq:q8}
\end{align}

\section{Green's function and SUSY partner potentials}

The Green's function defined in ({\ref{eq:g1}) and the associated sum rule is a special case, with $\epsilon=0$, of
\begin{align}
G(x,x^{\prime};\epsilon) &= \sum_j \frac{\psi_j(x)\psi_j(x^{\prime})}{(\epsilon - \gamma_j^2)} \label{eq:g21}\\
\Big[\frac{d^2}{dx^2} - U + \epsilon \Big]G(x,x^{\prime}) &=\delta(x-x^{\prime}) \label{eq:g22}\\
\int_a^b G(x,x;\epsilon) dx&= \sum_j \frac{1}{(\epsilon - \gamma_j^2)} \label{eq:g23}
\end{align}
The generalized Green's function may be constructed from two linearly independent solutions which satisfy boundary conditions at $x=a$ and $x=b$ by the following procedure:
\begin{align}
0&=\Big[\frac{d^2}{dx^2} - U + \epsilon \Big]\psi(x) \label{}\\
\psi^{\prime}(x)&=\psi(x)\int_b^x\frac{dy}{\psi^2(y)},\  \Rightarrow G(x,x^{\prime};\epsilon)= \psi(x) \psi^{\prime}(x^{\prime}) =\psi(x)\psi(x^{\prime})\int_b^{x>}\frac{dz}{\psi^2(z)}\label{eq:g24}
\end{align}
in which $\psi(x)$ satisfies a boundary condition at $x=a$ and $\psi^{\prime}(b)=0$, which is appropriate for studying sum rules for eigenstates satisfying the same boundary conditions. Another equivalent representation is
\begin{equation}
\psi(x)=\psi^{\prime}(x)\int_a^x\frac{dy}{\psi^{\prime2}(y)}\ \Rightarrow G(x,x^{\prime};\epsilon)= -\psi^{\prime}(x) \psi(x^{\prime}) =-\psi^{\prime}(x)\psi^{\prime}(x^{\prime})\int_a^{x<}\frac{dz}{\psi^{\prime2}(z)}\label{eq:g25}
\end{equation}
Using either of these representations of $G$ it is possible to examine a second order sum rule:
\begin{align}
\sum_j \frac{1}{(\epsilon - \gamma_j^2)^2} &= \int_a^b \int_a^b G(x,y;\epsilon) G(y,x;\epsilon) dx dy \label{}\\
&=\int_a^b dx \int_a^x dy\Big(\psi(x)\psi(y)\int_b^x\frac{dz}{\psi^2(z)}\Big)^2 \notag\\
&+\int_a^b dx \int_x^b dy\Big(\psi(x)\psi(y)\int_b^y\frac{dz}{\psi^2(z)}\Big)^2 \notag\\
&=2\int_a^b dx \Big(\psi(x)\int_b^x\frac{dz}{\psi^2(z)}\Big)^2 \Big(\int_a^x dy\ \psi^2(y)\Big)\notag\\
&=2\int_a^b dx \Big(\psi^2(x)\int_b^x\frac{dz}{\psi^2(z)}\Big)^2 \Big(\frac{1}{\psi^2(x)}\int_a^x dy \ \psi^2(y)\Big)\notag\\
&= -2\int_a^b  dx G^2(x,x;\epsilon) G^{\prime}(x,x;\epsilon) \label{eq:g26}
\end{align}
in which
\begin{equation}
 G^{\prime}(x,x;\epsilon)= -(\phi^{\prime})^2(x)\int_a^x \frac{dy}{(\phi^{\prime})^2(y)} \ \hbox{where}\ \phi^{\prime}(x)=\frac{1}{\psi(x)} \label{eq:g27}
\end{equation}
can be identified using (\ref{eq:g25}) and interpreted using Supersymmetric Quantum Mechanics (SUSYQM) (Witten 1981, Andrianov {\it et al} 1984, Sukumar 1985}). Using SUSYQM it can be recognized that if $\psi(x)$ is a nodeless, but unnormalizable solution at an energy $\epsilon$ below the ground state energy of $U(x)$ then $\phi^{\prime}(x)$ is a solution at energy $\epsilon$ for the SUSY partner potential
\begin{equation} 
U^{\prime}(x)= U(x) -2\frac{d^2}{dx^2}\ln[\psi(x)] \label{}
\end{equation}
If $\phi^{\prime}(x)$ is normalizable in $[a,b]$ then it can be the ground state eigenfunction for $U^{\prime}(x)$ and the eigenvalue spectrum for $U^{\prime}$ has $\epsilon$ as the lowest eigenvalue plus all the eigenvalues for $U$. If $\phi^{\prime}(x)$ is not normalizable then $U^{\prime}$ and $U$ have identical spectra and a Green's function for $U^{\prime}$ can be constructed in the form given in (\ref{eq:g21}). Thus we can conclude that the second order sum rule can be interpreted in terms of the equi-position Green's functions for $U$ and its SUSY partner $U^{\prime}$ evaluated for energy $\epsilon$ which is below the lowest eigenvalue for $U$.

\subsection{Example 1- Particle in a box}
We consider the sum rules for the eigenvalues for a particle in a box with eigenfunctions which go to zero at $x=0$ and $x=1$ and choose $\epsilon=0$. The solution $\psi(x)=x$
at zero energy gives $\phi^{\prime}(x)=1/x$. $\phi^{\prime}(x)$ is not normalizable in the domain $[0,1]$. Hence the SUSY partner and its spectrum, which is identical to that of $U$ are given by
\begin{equation}
U^{\prime}=U-2(d^2\ln(x)/dx^2)=U+ x^{-2},\quad \gamma_j=j^2\pi^2,\ j=1,2,\ \dots \label{}\\
\end{equation}
A result from SUSY Quantum Mechanics relates the eigenstates of SUSY partner potentials in the form
\begin{align}
\psi^{\prime}_j&= \frac{1}{\sqrt{\gamma_j^2-\epsilon}}\Big[\frac{d\psi_j}{dx} -\frac{d\ln[\psi(x)]}{dx} \psi_j\Big] \ \Rightarrow \label{}\\
\psi_j&=\sqrt{2} \sin(j\pi x),\ \psi(x)=x\ \Rightarrow \psi^{\prime}_j=\sqrt{2}\Big[\cos(j\pi x) - \frac{\sin(j\pi x)}{(j\pi x)}\Big] \label{eq:s1}
\end{align}
The eigenstates $\psi^{\prime}_j$ may be identified as $(j\pi x)$ times the spherical Bessel functions of order $1$ with argument $j\pi x$ (Abramowitz ans Stegun 1965 p438). Using 
\begin{align}
&\int_0^1 \Big(\frac{\sin(j\pi x) \sin(k\pi x)}{jk\pi^2 x^2} - \frac{j\pi\cos(j\pi x) \sin(k\pi x)+k\pi\sin(j\pi x) \cos(k\pi x)}{jk\pi^2 x}\Big)\ dx \label{}\\
&=-\int_0^1 \frac{d}{dx}\Big(\frac{\sin(j\pi x) \sin(k\pi x)}{jk\pi^2 x}\Big)=\frac{\sin(j\pi x) \sin(k\pi x)}{jk\pi^2 x} \vert_1^{0} =0\label{}
\end{align}
it can be verified that
\begin{equation}
\int_0^1 \psi^{\prime}_j(x)  \psi^{\prime}_k(x) \ dx = \delta_{jk} \label{}
\end{equation}
showing that the eigenfunctions $\psi^{\prime}_j(x)$ form an orthonormal set of functions and form a complete set for functions which satisfy $[\psi^{\prime}_j(0)]=0$ and $[\psi^{\prime}_j(1)]^2= 2$.

The Green's functions and the resulting first order sum rules for the eigenvalues are
\begin{align}
G(x,x^{\prime})&=-x x^{\prime}\int_1^{x_>}\frac{dy}{y^2}=x_<(x_>-1)\ \Rightarrow \int_0^1 G(x,x)\  dx = \frac{-1}{6}= -\sum_j\frac{1}{j^2\pi^2}\label{}\\
G^{\prime}(x,x^{\prime})&=-\frac{1}{xx^{\prime}}\int_0^{x_<} z^2\ dz= -\frac{x_<^2}{3x_>}\  \Rightarrow \int_0^1 G^{\prime}(x,x)\ dx = \frac{-1}{6} =-\sum_j\frac{1}{j^2\pi^2}\label{}
\end{align}
The second order sum rules may be evaluated in 4 different ways : either by using $G(x,x^{\prime})$ in a two-dimensional integral or by using $G^{\prime}(x,x^{\prime})$ in a two-dimensional integral or using both $G(x,x)$ and $G^{\prime}(x,x)$ in two different one-dimensional integral as shown below:
\begin{align}
S_2&\equiv\sum_j \frac{1}{j^4\pi^4}\label{}\\
S_2&=\int_0^1 \int_0^1 G(x,y) G(y,x)\ dx\ dy= \int_0^1\Big[(1-x)^2\int_0^x y^2\ dy+x^2 \int_x^1 (1-y)^2\ dy\Big] \notag\\
&= \int_0^1\frac{x^2(1-x)^2}{3}\ dx=\frac{1}{90}\label{}\\
S_2&=\int_0^1 \int_0^1 G^{\prime}(x,y) G^{\prime}(y,x)\ dx\ dy= \frac{1}{9}\int_0^1\Big[x^{-2}\int_0^x y^4\ dy+x^4 \int_x^1 \frac{dy}{y^2}\Big]\ dx \notag\\
&= \frac{1}{9}\int_0^1\Big[\frac{x^3}{5} -x^4+x^3\Big]\ dx =\frac{1}{90}\label{}\\
S_2&=-2\int_0^1 G^2(x,x)\ G^{\prime}(x,x)\ dx= 2\int_0^1 x^2(1-x)^2 \frac{x}{3}\ dx = \frac{1}{90} \label{}\\
S_2&=-2\int_0^1 G^{\prime2}(x,x)\ G(x,x)\ dx= 2\int_0^1 \frac{x^2}{9} x(1-x)\ dx = \frac{1}{90} \label{}
\end{align}
in agreement with well  known results.

\subsection{Example 2- Simple harmonic Oscillator}

Consider the Hamiltonian for a harmonic oscillator in units in which $\hbar=1$, $m=1/2$ and $\omega=2$ so that the Schrodinger equation is
\begin{equation}
\frac{d^2\psi}{dx^2}= (x^2-E)\psi \label{}
\end{equation}
The minimum value of the potential is $0$ at $x=0$, the energy spacing is $2$, $E_n=2n+1$ and the ground state is at $E=1$. If we consider the even states which are symmetric in $x$ and are orthogonal in the space $[0,\infty]$ and are spaced in units of $4$ they can be denoted by $\psi_{2n}$ with energy $E_{2n}=(4n+1)$. We can construct a Green's function for the sum rules for the even states by the following procedure. A solution at energy $E=-1$ which is below the minimum of the potential is $\psi=\exp(+0.5 x^2)$ which is not normalizable but $\phi=(\psi)^{-1}= \exp(-0.5 x^2)$ is a normalizable function and can be the ground state eigenfunction for a SUSY partner potential. Hence
\begin{equation}
U^{\prime}=U- 2\frac{d^2\ln(\psi)}{dx^2} =x^2-2 \label{}\\
\end{equation}
which is an oscillator potential shifted downwards by $2$ and as a spectrum which has all the eigenvalues of $U$ and in addition has a ground state at $E=-1$ and the spectrum can be identified as $E^{\prime}_n= 2n-1, n=0,1,2,\ \dots$. The first order sum rule for the eigenvalues of the oscillator does not converge but diverges logarithmically. For the second order sum rule the Green's functions for $U$ and $U^{\prime}$ may be constructed using the solutions for $E=-1$.

Since $\psi(x)$, the solution for $E=-1$, has vanishing first derivative at $x=0$, a sum rule for the symmetric eigenstates states which also have vanishing first derivatives at $x=0$ may be constructed 
\begin{align}
\psi&=\exp\Big(\frac{x^2}{2}\Big),\ \psi^{\prime}=\exp\Big(\frac{x^2}{2}\Big) \int_{\infty}^x \exp(-z^2)\ dz,\ \Rightarrow \label{}\\
G(x,x)&=- \exp(x^2)\int_{\infty}^{x} \exp(-z^2)\ dz \label{}\\
\phi&=\exp\Big(-\frac{x^2}{2}\Big),\ \phi^{\prime}=\exp\Big(-\frac{x^2}{2}\Big) \int_{0}^x \exp(y^2)\ dy\ \Rightarrow\label{}\\
G^{\prime}(x,x)&=\exp(-x^2)\int_{0}^{x} \exp(y^2)\ dy \label{}\\
S_2&=2\int_0^{\infty} G^2(x,x) G^{\prime}(x,x)\ dx \notag\\
&=2\int_0^{\infty}\Big(\Big[\exp(x^2)\int_x^{\infty} \exp(-z^2)\ dz\Big]^2\ \exp(-x^2)\int_{0}^{x} \exp(y^2)\ dy \Big) \label{eq:SS1}\\
S_2&=2\int_0^{\infty} G^{\prime2}(x,x) G(x,x)\ dx \notag\\
&=2\int_0^{\infty}\Big(\Big[\exp(-x^2)\int_{0}^{x} \exp(y^2)\ dy\Big]^2\ \exp(x^2)\int_x^{\infty} \exp(-z^2)\ dz \Big) \label{eq:SS2}\\
S_2&=\sum_{j=0}^{\infty} \frac{1}{(E_{2j}-E^{\prime}_0)^2} = \sum_{j=0}^{\infty} \frac{1}{(4j+2)^2} = \frac{\pi^2}{32} \label{eq:SS3}
\end{align}
where the sum is over the eigenvalues of the symmetric states of $U$. The equality of the expressions in (\ref{eq:SS1}), (\ref{eq:SS2}) and (\ref{eq:SS3}) has been verified by performing the integrations involving the Dawson's integral and the Complementary Error function (Abramowitz and Stegun 1965 p297-8).

\subsection{Power law Confining Potentials}

The procedure discussed for the addition of a state and finding sum rule for eigenvalues may be extended for all confining potentials of the power law form. We consider potentials of the form $U=|x|^n$ which support an infinite number bound states with positive definite energies and examine the  Green's function for finding sum rules for the eigenstates of $U$ and also finding a SUSY partner which has an additional bound state at $E=0$ :
\begin{align}
\frac{d^2\psi}{dx^2}&=\vert x\vert^{\alpha} \psi,\  \nu\equiv \frac{1}{\alpha+2},\ z\equiv 2\nu x^{\frac{1}{2\nu}} \Rightarrow \Big(\frac{d^2}{dz^2}+\frac{1}{z}\frac{d}{dz}- \big[1+\frac{\nu^2}{z^2}\Big)\psi=0 \label{}\\
\psi^{(1)}&= \sqrt{x} K_{\nu}(2\nu x^{\frac{1}{2\nu}}) ,\ \psi_{\pm}^{(2)} =\sqrt{x} I_{\pm\nu}(2\nu x^{\frac{1}{2\nu}})\ \Rightarrow \psi_-^{(2)}=\psi_+^{(2)}+\frac{2}{\pi}\sin(\pi\nu) \psi^{(1)} \label{}
\end{align}
where $K_{\nu}$ and $I_{\pm\nu}$ are modified Bessel functions (Abramowitz and Stegun 1965 p374-380). The limiting values of the different solutions at $x=0$ and for large absolute values of $x$ are: 
\begin{align}
\psi^{(1)}(x)\vert_0 &\sim x^0,\ \psi_{+}^{(2)}(x)\vert_0 \sim x,\ \psi_{-}^{(2)}\vert_0\sim x^0, 
\ \frac{d\psi_{-}^{(2)}}{dx}\vert_0=0\ \label{}\\
\psi_{\pm}^{(2)}&\vert_{\infty}\sim\sqrt{\frac{x}{z}}\exp(z)=x^{\frac{\alpha}{4}}\exp(z),\ \psi^{(1)}(x)\vert_{\infty} \sim \sqrt{\frac{x}{z}}\exp(-z)=x^{\frac{\alpha}{4}}\exp(-z)\label{}
\end{align}
These limiting values suggest appropriate functions for the construction of Green's function for the symmetric states which have vanishing derivatives at $x=0$ and have vanishing values in the asymptotic region $x^2 \rightarrow  \infty$ is
\begin{equation}
G_e(x,x^{\prime})= -2\nu\sqrt{xx^{\prime}} I_{-\nu}(2\nu x^{\frac{1}{2\nu}})\ K_{\nu}(2\nu x^{\prime\frac{1}{2\nu}}) \label{eq:pg1}\\
\end{equation}
The Green's function appropriate for the antisymmetric states which vanish at $x=0$ and have vanishing values in the asymptotic region is
\begin{equation}
G_o(x,x^{\prime})=- 2\nu\sqrt{xx^{\prime}} I_{+\nu}(2\nu x^{\frac{1}{2\nu}})\ K_{\nu}(2\nu x^{\prime\frac{1}{2\nu}}) \label{eq:pg2}\\
\end{equation}
The WKB result for the eigenvalues $E_n$ of $U$ in the limit of large quantum number $n$ can be derived from the Bohs-Sommerfield formula and is of the form
\begin{equation}
E_n=\gamma_n^2\sim \Big(\Big[n+\frac{1}{2}\Big]\frac{\sqrt{\pi}(\alpha+2)\Gamma\big(\frac{\alpha+2}{\alpha}\big)}{2\Gamma\big(\frac{1}{\alpha}\big)}\Big)^{\frac{2\alpha}{\alpha+2}} \label{}
\end{equation}
which shows that the sums over the inverse of eigenvalues will not converge if $\alpha<2$. For $\alpha>2$, (i.e) $\beta<1/4$,  the Green's functions in (\ref{eq:pg1}) and (\ref{eq:pg2}) may be used in the integration range $[0,\infty]$ and the integrals may be performed (Gardshteyn and Ryzhik p 693-4) to find the sum rules
\begin{align}
S_{even}&=\sum_{n=0}^{\infty}\frac{1}{E_{2n}} = -\int_0^{\infty} G_o(x,x) dx=\nu^{2-4\nu}\frac{\Gamma(2\nu)\Gamma(\nu)\Gamma(1-4\nu)}{\Gamma(1-3\nu)\Gamma(1-2\nu)},\ 0<\nu<\frac{1}{4} \label{eq:pg3}\\
S_{odd}&=\sum_{n=0}^{\infty}\frac{1}{E_{2n+1}} = -\int_0^{\infty} G_e(x,x) dx=\nu^{2-4\nu}\frac{\Gamma(3\nu)\Gamma(2\nu)\Gamma(1-4\nu)}{\Gamma(1-2\nu)\Gamma(1-\nu)},\ 0<\nu<\frac{1}{4} \label{eq:pg4}\\
\end{align}
The WKB formula for the eigenvalues in the large $n$ limit also show that the difference between the sums over the inverses of the even and odd eigenvalues will converge for positive values of $\alpha$. It can be shown that
\begin{align}
&S=\sum_{n=0}^{\infty}\frac{(-)^n}{E_{n}} = -\int_0^{\infty}\big( G_e(x,x) - G_o(x,x\big) dx =\nu^{2-4\nu}\frac{\Gamma(3\nu)\Gamma^2(2\nu)}{\Gamma(4\nu)\Gamma(1-\nu)}, 0<\nu<\frac{1}{2} \label{eq:pg5}\\
&S_{even}= S\ \frac{\sin(3\pi\nu)}{\sin(\pi\nu)}\ \frac{1}{2\cos(2\pi\nu}),\ 0<\nu<\frac{1}{4}\label{}\\
&S_{odd}= S\ \frac{1}{2\cos(2\pi\nu)},\ 0<\nu<\frac{1}{4} \label{}
\end{align}

It is also possible to use the zero energy solutions in $U(x)$ to find a SUSY partner $U^{\prime}(x)$ which has a spectrum with all the eigenvalues of $U$ and in addition has an extra boundstate at $E=0$. It can be shown that $\Psi=\psi_{-}^{(2)}$ is a symmetric nodeless function which is not normalizable but $\Phi=1/\Psi$ is normalizable and can be the ground state eigenfunction for the SUSY partner potential
\begin{equation}
\Psi \equiv \sqrt{x} I_{-\nu}(2\nu x^{\frac{1}{2\nu}}) ,\quad U^{\prime}=\vert x\vert^n -2\frac{d^2\ln \Psi}{dx^2} \label{}
\end{equation}
Using the properties of the Modified Bessel functions it can be shown that
\begin{align}
U^{\prime}(x)&=\vert x\vert^n\ \Big(-1+2\Big[\frac{I_{1-\nu}(2\nu x^{\frac{1}{2\nu}})}{I_{-\nu}(2\nu x^{\frac{1}{2\nu}})}\Big]^2\Big) \label{}\\
x \rightarrow &0,\ U^{\prime}(x) \rightarrow -\vert x\vert^n\ +\frac{2\vert x\vert^{2n+2}}{(n+1)^2}\  ;\quad x \rightarrow \infty,\  U^{\prime}(x) \rightarrow
\vert x\vert^n -n \vert x\vert^{\frac{n-2}{2}}+O(x^{-2})\label{}
\end{align}
The SUSY partner potentials with an extra boundstate at $x=0$ for $n=2,\ 4,\ 6$ and $8$ are shown in figures $1$ and $2$.
\begin{figure}[htb]
\centering
\includegraphics[width=7cm]{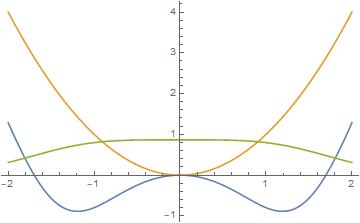}\ \includegraphics[width=7cm]{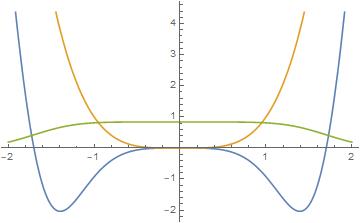}
\caption{{\small $U= x^n$ (yellow) and $U^{\prime}$ (blue) in the $x$ range $[-2,2] $ for $n=2$ and $n=4$}.  The unnormalized eigenfunction for $U^{\prime}$ at $E=0$ is shown in green. }
\end{figure}
\begin{figure}[htb]
\centering\
\includegraphics[width=7cm]{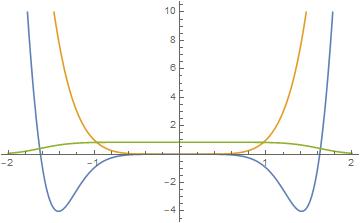}\ \includegraphics[width=7cm]{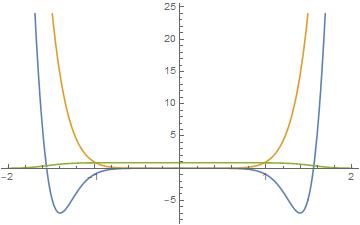}
\caption{{\small $U= x^n$ (yellow) and $U^{\prime}$ (blue) in the $x$ range $[-2,2] $ for $n=6$ and $n=8$}. The unnormalized eigenfunction for $U^{\prime}$ at $E=0$ is shown in green.}
\end{figure}

The pattern of change of $U^{\prime}$ as $n$ increases is evident. $U^{\prime}$ is of the form of a double well with a flat hump and the groundstate energy level grazes the flat hump. The double wells are located close to $x^2=1$ and the flat region widens as n increases. In the limit of large $n$ both $U$ and $U^{\prime}$ become very large as $x^2$ exceeds $1$ and the pockets of the double wells located near $x^2=1$ become narrower and deeper as n increases. In the limit of $n\rightarrow \infty$ both $U$ and $U^{\prime}$ approach infinite walls as $x^2$ exceeds $1$ and $U^{\prime}$ has deep and narrow wells when $x^2$ is just below $1$. For large values of $n$ the groundstate wave function for $U^{\prime}$ will be  almost a constant in the flat region $x^2\le 1-\epsilon$, $\epsilon<<1$, and dip smoothly to zero close to $x^2\sim 1$ and will look almost like a square barrier of height $\sim 1/\sqrt{2}$ and width $\sim 2$ confined to the region $x^2<1$.

A Green's function can be constructed in the form
\begin{equation}
G^{\prime}(x,x^{\prime}) = -\frac{1}{\Psi(x)\Psi(x^{\prime})}\int_0^{x_<}\Psi^2(y)\ dy \label{}
\end{equation}
leading to a second representation of the first order sum rule for the symmetric eigenstates of $U$ of the form
\begin{align}
\sum_{n=0}^{\infty} \frac{1}{E_{2n}}&=-\int_0^{\infty} G^{\prime}(x,x)\ dx=\int_0^{\infty}\frac{dx}{xI^2_{-\nu}(2\nu x^{\frac{1}{2\nu}})}\int_0^{x}yI^2_{-\nu}(2\nu y^{\frac{1}{2\nu}})\ dy \label{}\\
&=\big(2\nu\big)^{2-4\nu}\int_0^{\infty}\frac{dz}{z}I^{-2}_{-\nu}(z)\int_0^z y^{4\nu-1} I^{2}_{-\nu}(y) dy \label{}
\end{align}
which converges, for $n>2$, to the value given in (\ref{eq:pg3}). This identification leads to the relation
\begin{align}
\int_0^{\infty} z^{4\nu-1} I_{-\nu}(z)\ K_{\nu}(z) dz &=
\int_0^{\infty}\frac{dz}{z}I^{-2}_{-\nu}(z)\int_0^{z} y^{4\nu-1} I^2_{-\nu}(y)\ dy \label{}\\
&=2^{4\nu-2}\frac{\Gamma(2\nu)\Gamma(\nu)\Gamma(1-4\nu)}{\Gamma(1-3\nu)\Gamma(1-2\nu)},\ 0<\nu<\frac{1}{4} \label{}
\end{align}

A similar analysis of the sum rules for the odd eigenstates can be shown to lead to the relation
\begin{align}
\int_0^{\infty} z^{4\nu-1} I_{\nu}(z)\ K_{\nu}(z) dz &=
\int_0^{\infty}\frac{dz}{z}I^{-2}_{\nu}(z)\int_0^{z} y^{4\nu-1} I^2_{\nu}(y)\ dy \label{}\\
&=2^{4\nu-2}\frac{\Gamma(3\nu)\Gamma(2\nu)\Gamma(1-4\nu)}{\Gamma(1-2\nu)\Gamma(1-\nu)},\ 0<\nu<\frac{1}{4} \label{}
\end{align}
These relations arise because it can be shown using the differential equation satisfied by the Modified Bessel functions that
\begin{equation}
K_{\nu}(z)=I_{\pm \nu}(z)   \int_z^{\infty} \frac{dy}{y} I^{-2}_{\pm \nu}(y) \label{}
\end{equation}

\section{Conclusion}

It has been shown that Green's functions for reflectionless potentials lead to a hierarchy of sumrules. The methods of Supersymmetric Quqantum Mechanics have been used to study the relation between Green's functions for SUSY partners. Potentials with an extra zero energy boundstate in addition to the energy spectrum of powerlaw potentials have been constructed and analyzed. For the case of a free particle confined to the space $[0,1]$, by considering the solutions at energy $E=0$, a SUSY partner potential can be identified for which the eigenfunctions are shown to be identical to the solutions of the radial Schroedinger equation for $l=1$ for a free particle which are of the form $R(kr)\sim krj_1(kr)$ with $k=n\pi,\ n=1,2,\dots$\ . Hence the set of functions $R(n\pi r)$ form a set of orthonormal functions in the interval $[0,1]$. This identification provides an alternative basis for expanding functions in the interval $[0,1]$ instead of the usual Fourier series basis set.

\section*{References}

\noindent Abramowitz M.and Stegun I.A. 1965 {\it{Handbook of Mathematical Functions}} Dover Publications  297, 374, 438

\noindent A.C.Scott, F.Y.E.Chu and D.W.Mclaughlin, Proc. I.E.E.E. {\bf 61}, 1443 (1973)

\noindent Andrianov A A, Borisov N V and Ioffe M V 1984 {\it Phys. Lett.} {\bf 105A} 19

\noindent C.S.Gardner, J.M.Greene, M.D.Kruskal and R.M.Miura, Phys. Rev. Lett. {\bf 19}, 1095 (1967)

\noindent Gradshteyn I.S. and Ryzhik I.M. 1965 {\it{Table of Integrals, Series and Products}} Academic Press 38

\noindent P.D.Lax, Comm. Pure. Appl. Math. {\bf 21}, 467 (1968)

\noindent Sukumar C V 1985a {\it J. Phys. A: Math. Gen.} {\bf 18} 2917

\noindent Sukumar C V 1985b {\it J. Phys. A: Math. Gen.} {\bf 18} 2937

\noindent Witten E 1981 {\it Nucl. Phys. B} {\bf 188} 513

\end{document}